\documentclass[a4paper,12pt]{article}

\pdfoutput=1
\usepackage{amsmath}
\usepackage{amssymb}
\usepackage{amsfonts}
\usepackage{mathrsfs}
\usepackage{bbm}
\usepackage{graphicx,subfigure,booktabs}
\usepackage[numbers,sort&compress]{natbib}
\usepackage{verbatim}
\usepackage{color}
\usepackage{ulem}
\usepackage{setspace}
\usepackage{multirow}
\usepackage{url}

\usepackage[utf8]{inputenc}

\usepackage[colorlinks,
linkcolor=black,
filecolor=black,
anchorcolor=black,
urlcolor=black,
citecolor=blue,
bookmarks=false,
]{hyperref}

\usepackage[hyphenbreaks]{breakurl}

\usepackage{fancyhdr}

\numberwithin{equation}{section}

\newlength{\dinwidth}
\newlength{\dinmargin}
\allowdisplaybreaks

\begin{document}



\title{Explaining the $B_{d(s)} \rightarrow K^{(\ast)}\bar{K}^{(\ast)}$ puzzle via chiral-flip in $R$-parity violating MSSM with seesaw mechanism}

\author{
Min-Di Zheng${}^{1}$\footnote{zhengmd5@mail.sysu.edu.cn}\,\,,
Qi-Liang Wang${}^{1}$\footnote{wangqiliang2024@163.com}\,\,,
Li-Fen Lai${}^{1}$\footnote{lailifen@mails.ccnu.edu.cn}\,\,,
and 
Hong-Hao Zhang${}^{2}$\footnote{zhh98@mail.sysu.edu.cn}\\[12pt]
\small ${}^{1}$ School of Physical Science and Intelligent Education,\\ [-0.2cm]
\small Shangrao Normal University, Shangrao 334001, China \\[-0.2cm]
\small ${}^{2}$ School of Physics, Sun Yat-Sen University, Guangzhou 510275, China
}

\date{}
\maketitle

\begin{abstract}
We study the non-leptonic puzzle of $B_{d(s)} \rightarrow K^{(\ast)}\bar{K}^{(\ast)}$ decay in the $R$-parity violating minimal supersymmetric standard model (RPV-MSSM) extended with the inverse seesaw mechanism. In this model, the chiral flip of sneutrinos can contribute to the observables $L_{K\bar{K}}$ and $L_{K^{\ast}\bar{K}^{\ast}}$, that is benefit for explaining the relevant puzzle. We also find that this unique effect can engage in the $B_s$-$\bar{B}_s$ mixing. We utilize the scenario of complex $\lambda^\prime$ couplings to fulfill the recent stringent constraint of $B_s$-$\bar{B}_s$ mixing, and examine other related bounds of $B,K$-meson decays, lepton decays, neutrino data, $Z$ decays, oblique parameters, CP violations (CPV), etc. Besides, inspired by the new measurement of ${\cal B}(B^+ \rightarrow K^+\nu\bar{\nu})$ by Belle II, which shows about $2.7\sigma$ higher than the Standard Model (SM) prediction, we also investigate the New Physics (NP) enhancement to this observable.
\end{abstract}
\newpage

\section{Introduction}\label{sec:intro}
In recent years, series of deviations between the experimental measurements and the SM predictions have been witnessed in the context of $B$-meson semileptonic decays, e.g. the lepton flavor universality (LFU) ratios $R_{D^{(\ast)}}$. However, another type of LFU ratios, $R_{K^{(\ast)}}$ within the $b\to s\ell^+\ell^-$ ($\ell=e,\mu$) processes, has recently been reported in agreement with SM predictions~\cite{LHCb:2022qnv}, and it is already erased from the anomaly list. Since the LFU violation from the NP still needs time to be confirmed, there exist U-spin related observables within rare $b\rightarrow s(d)$ transitions, i.e. $L_{K^{(\ast)}\bar{K}^{(\ast)}}$~\cite{Biswas:2023pyw}, which can also be utilized to search for NP clues. The observables $L_{K^{(\ast)}\bar{K}^{(\ast)}}$, defined as the ratios of longitudinal branching ratios ($\bar{B}^{(\ast)}_s \rightarrow K^{(\ast)}\bar{K}^{(\ast)}$ versus $\bar{B}^{(\ast)}_d \rightarrow K^{(\ast)}\bar{K}^{(\ast)}$), are recently measured~\cite{ParticleDataGroup:2022pth,BaBar:2007wwj,LHCb:2019bnl,BaBar:2006enb,Belle:2012dmz,Belle:2015gho,LHCb:2020wrt}:
\begin{align}
L^{\rm exp}_{K^{\ast}\bar{K}^{\ast}}=4.43 \pm 0.92,  \quad\quad
L^{\rm exp}_{K\bar{K}}=14.58 \pm 3.37,
\end{align} 
showing the $2.6\sigma$($2.4\sigma$) pull values corresponding to the SM predictions within QCD factorisation~\cite{Biswas:2023pyw}:
\begin{align}
L^{\rm SM}_{K^{\ast}\bar{K}^{\ast}}=19.53^{+9.14}_{-6.64},  \quad\quad
L^{\rm SM}_{K\bar{K}}=26.00^{+3.88}_{-3.59}.
\end{align} 
This puzzle implies that there may exist new quark-flavor structure in NP. For the model-independent discussion, the Lagrangian of the low energy effective field theory is given by
\begin{align}
{\cal L}_{\rm eff} = \frac{4 G_F}{\sqrt{2}} \eta_t \sum_{i}C_i{\cal O}_i + {\rm h.c.},
\end{align}
where the Cabibbo–Kobayashi–Maskawa (CKM) factor $\eta_t \equiv K_{tb} K_{tp}^{\ast}$~($p=s,d$). The most relevant operators for the puzzle-explanation are the given QCD penguin operators and magnetic operators~\cite{Beneke:2001ev}, 
\begin{align}
{\cal O}_{4 p}&=(\bar{p}_L^\alpha \gamma^\mu  b_L^\beta)
\sum_q (\bar{q}_L^\beta \gamma_\mu  q_L^\alpha), &\qquad
{\cal O}_{6 p}&=(\bar{p}_L^\alpha \gamma^\mu  b_L^\beta)
\sum_q (\bar{q}_R^\beta \gamma_\mu  q_R^\alpha),  \notag \\
{\cal O}_{7\gamma p}&=\frac{-e m_b}{16\pi^2} (\bar{p}_L^\alpha \sigma^{\mu\nu} b_R^\alpha) F_{\mu\nu}, &\qquad
{\cal O}_{8g p}&=\frac{-g_s m_b}{16\pi^2} (\bar{p}_L^\alpha \sigma^{\mu\nu} T^a_{\alpha\beta} b_R^\beta) G^a_{\mu\nu},
\end{align}
where $\alpha,\beta$ are color indices and a summation over $q=u,d,c,s,b$ is implied, with the vertex couplings $+ig_s T^a$ and $+iQ_e e$ for $Q_e=-1$. The recent global fit results~\cite{Alguero:2020xca,Biswas:2023pyw,Biswas:2024bhn} show that, for the $1\sigma$ level, ones need the negative $C_{8gs}^{\rm NP}$ (positive $C_{8gd}^{\rm NP}$) with the value of ${\cal O}(10^{-1})$, or positive $C_{4s}^{\rm NP}$ (negative $C_{4d}^{\rm NP}$) with the value of ${\cal O}(10^{-2})$, while single positive $C_{6s}^{\rm NP}$ around ${\cal O}(10^{-2})$ can only explain the tension of $L_{K\bar{K}}$.

Since the recent model-independent researches throw light on the regions of Wilson coefficients, in this work, we will investigate this puzzle in a concrete NP model. Inspired by the recent research on the gluon-penguin contributions, within the $S_1$-leptoquark model containing the $U(q)_{1,2}$  flavor symmetry and inverse seesaw mechanism~\cite{Lizana:2023kei}, we utilize the RPV-MSSM extended with the inverse seesaw mechanism (named as RPV-MSSMIS). It is worth mentioning that we had recently proposed this model to study LFU observables, i.e. $R_{K^{(\ast)}}$ and $R_{D^{(\ast)}}$, as well as the muon anomalous magnetic moment~\cite{Zheng:2021wnu,Zheng:2022ssr}, and this model can provide the particular feature for different  quark flavor through the $\lambda'$-coupling texture. In this work, we find that the chiral flip of sneutrino can make unique contributions to the Wilson coefficients $C_{8gs(d)}^{\rm NP}$,  extracted from the gluon-penguin diagrams, through mainly the $\tilde{\nu}dd$ loop.  Also, the strict constraint from the rare decay $B \rightarrow X_s \gamma$, can be relaxed by the cancellation of $C_{8gs(d)}^{\rm NP} = -3 C_{7 \gamma s(d)}^{\rm NP}$ (this relation is induced by the model feature). In this work, we scrutinize all the one-loop gluon($\gamma$)-penguin diagrams of $b$ interaction to $s(d)$, as well as the calculations in other related processes, within the RPV-MSSMIS. Among these, we also find significant chiral-flip contribution to the $B_s$-$\bar{B}_s$ mixing, and this effect on $B\rightarrow K^{(\ast)} \nu\bar{\nu}$ decays. Recently, Belle II Collaboration has reported the new measurement of the branching ratio, ${\cal B}(B^+ \rightarrow K^+ \nu \bar{\nu})_{\rm exp}=(2.3\pm 0.7) \times 10^{-5}$~\cite{Belle-II:2023esi}, higher than the corresponding SM prediction~\cite{Becirevic:2023aov} by around $2.7\sigma$. As is known to all, this $b$ decaying into $s$ mode is one of the cleanest probes for NP searches due to its highly suppressed theoretical uncertainty. Here we revisit the NP contributions to the $b\rightarrow s \nu \bar{\nu}$ transition and discuss the enhancement effects. 

This paper is organized as follows. The RPV-MSSMIS model and the theoretical calculations are in Sec.~\ref{sec:RPV-MSSMIS}. Then, in Sec.~\ref{sec:constraints}, we scrutinize the related constraints, which are followed by numerical results and discussions in Sec.~\ref{sec:num} and additional discussions on CPV in Sec.~\ref{sec:addition}. Our conclusions are presented in Sec.~\ref{sec:conclusion}.   

\section{The tension study in RPV-MSSMIS}\label{sec:RPV-MSSMIS}

In this section, the NP effects, especially the chiral-flip ones, are investigated in the $B_{d(s)} \rightarrow K^{(\ast)}\bar{K}^{(\ast)}$ and $B \rightarrow K^{(\ast)} \nu\bar{\nu}$ decays, within the RPV-MSSMIS. 

\subsection{RPV-MSSMIS framework}
\label{sec:model}
First let us briefly review the RPV-MSSMIS~\cite{Zheng:2021wnu}. Here are given the superpotential and the soft supersymmetric (SUSY) breaking Lagrangian,
\begin{align}\label{eq:MSSMIS-RPV}
{\cal W} =& {\cal W}_{\rm MSSM} 
+ Y_\nu^{ij} \hat R_i \hat L_j \hat H_u + M_R^{ij} \hat R_i \hat S_j + \frac{1}{2} \mu_S^{ij} \hat S_i \hat S_j
+ \lambda'_{ijk} \hat L_i \hat Q_j \hat D_k,   \notag\\
{\cal L}^{\rm soft}=&{\cal L}^{\rm soft}_{\rm MSSM}
-(m^2_{\tilde{R}})_{ij} \tilde{R}^\ast_i\tilde{R}_j
-(m^2_{\tilde{S}})_{ij} \tilde{S}^\ast_i\tilde{S}_j \notag\\
&-(A_\nu Y_{\nu})_{ij} \tilde{R}^\ast_i\tilde{L}_jH_u
-B_{M_R}^{ij} \tilde{R}^\ast_i\tilde{S}_j
-\frac{1}{2} B_{\mu_S}^{ij} \tilde{S}_i\tilde{S}_j ,
\end{align} 
where the generation indices $i,j,k=1,2,3$ while the colour ones are omitted, and squarks (sleptons) are denoted by the symbol ``$\tilde{\ }$'', and as for the MSSM parts, ${\cal W}_{\rm MSSM}$ and ${\cal L}^{\rm soft}_{\rm MSSM}$, the reader can refer to Refs~\cite{Rosiek:1989rs,Rosiek:1995kg}. All repeated indices are assumed to be summed over throughout this paper unless otherwise stated. The neutral scalar fields of the two Higgs doublet superfields, $\hat H_u=(\hat H^+_u,\hat H^0_u)^T$ and $\hat H_d=(\hat H^0_d,\hat H^-_d)^T$, acquire the non-zero vacuum expectation value, i.e. $\langle H^0_u \rangle=v_u$ and $\langle H^0_d \rangle=v_d$, respectively, and their mixing is expressed by $\tan\beta=v_u/v_d$.

The neutrino sector in the superpotential ${\cal W}$ provides the neutrino mass spectrum at the tree level, and in the $(\nu, R, S)$ basis, the $9\times9$ mass matrix ${\cal M}_{\nu}$ is given by
\begin{align}\label{eq:mnu}
{\cal M}_{\nu} = \left( 
\begin{array}{ccc}
0 &m_D^{T}  &0\\ 
m_D  &0 &M_R\\ 
0 &M_{R}^{T} &\mu_S\end{array} 
\right), 
\end{align}
where the Dirac mass matrix $m_D = \frac{1}{ \sqrt{2}} v_u Y_\nu^T$. Then ones can diagonalize ${\cal M}_{\nu}$ through ${\cal M}^{\text{diag}}_{\nu}={\cal V} {\cal M}_{\nu} {\cal V}^T$. As to the sneutrino mass square matrix ${\cal M}_{\tilde{\nu}^{\cal I(R)}}^2$ in the $(\tilde{\nu}^{\cal I(R)}_{L},\tilde{R}^{\cal I(R)},\tilde{S}^{\cal I(R)})$ basis, it is expressed as
\begin{align}\label{eq:mSnu}
{\cal M}_{\tilde{\nu}^{\cal I(R)}}^2  =& 
\left(\begin{array}{ccc} 
m^2_{\tilde{L}'} & (A_\nu -\mu\cot\beta) m_D^T & m_D^T M_R \\
(A_\nu -\mu\cot\beta) m_D & m^2_{\tilde{R}}+M_RM_R^{T}+m_Dm_D^{T} & 
\pm M_R\mu_S + B_{M_R} \\
M_R^T m_D & \pm \mu_S M_R^{T} + B_{M_R}^T 
& m^2_{\tilde S}+ \mu_S^2+M_R^TM_R \pm B_{\mu_S}
\end{array}\right)  \notag\\
\approx &
\left(\begin{array}{ccc} 
m^2_{\tilde{L}'} & (A_\nu -\mu\cot\beta) m_D^T & m_D^T M_R \\
(A_\nu -\mu\cot\beta) m_D & m^2_{\tilde{R}}+M_RM_R^{T}+m_Dm_D^{T} & B_{M_R} \\
M_R^T m_D & B_{M_R}^T 
& m^2_{\tilde S}+M_R^T M_R \pm B_{\mu_S}
\end{array}\right),
\end{align}
where the ``$\pm$'', as well as ``${\cal R(I)}$'', denotes the even (odd) CP, and the mass square $m^2_{\tilde{L}'} \equiv m^2_{\tilde{L}}+\frac{1}{4} g^2_2 v^2 \cos 2\beta+m_Dm_D^T$ is regarded as the model input, with $m^2_{\tilde{L}}$ being the soft mass square of $\tilde{L}$. The parameter $\mu_S$ is generally tiny inducing the smallness of active neutrino mass~\cite{DeRomeri:2018pgm}, while $B_{\mu_S}$ is considered non-negligible, which induces the mass splitting between the CP-even and CP-odd sneutrinos for the same flavor. This is different from the degenerate-mass approximation adopted in our recent researches~\cite{Zheng:2021wnu,Zheng:2022ssr}\footnote{Although the quasi-degenerate-mass scenario is favored by the direct dark matter (DM) detection~\cite{An:2011uq}, we focus on the field of $B$-meson processes, and given RPV is involved, DM is out of the scope of this work.}. 
In the following sections, ones will see that this splitting-mass scenario can provide the chiral-flip contributions in some processes.

Afterwards we introduce the trilinear RPV interaction in this model. The superpotential term $\lambda'_{ijk} \hat L_i \hat Q_j \hat D_k$ induces the relevant Lagrangian in the context of mass eigenstates for the down-type quarks and charged leptons, which is given by
\begin{align}\label{eq:RPVlagphys}
{\cal L}_{\text{LQD}} =& \lambda'^{\cal I(R)}_{vjk} \tilde{\nu}_{v}^{\cal I(R)} \bar{d}_{Rk} d_{Lj} 
+\lambda'^{\cal N}_{vjk} \big(\tilde{d}_{Lj} \bar{d}_{Rk} \nu_{v} + \tilde{d}_{Rk}^\ast \bar{\nu}_{v}^c d_{Lj} \big)  \notag\\
&-\tilde{\lambda}'_{ilk} \big(\tilde{l}_{Li} \bar{d}_{Rk} u_{Ll} + \tilde{u}_{Ll} \bar{d}_{Rk} l_{Li} + \tilde{d}_{Rk}^\ast \bar{l}_{Li}^c u_{Ll}\big) + {\rm h.c.},
\end{align}
where ``$c$'' indicates the charge conjugated fermions, and the fields $\tilde{\nu}_{L}^{\cal I(R)}$, $\nu_{L}$, and $u_{L}$ (aligned with $\tilde{u}_L$) in the flavor basis have been rotated into mass eigenstates by the mixing matrices $\tilde{\cal V}^{\cal I(R)}$, ${\cal V}$, and $K$, respectively. Besides, the index $v=1,2,\dots 9$ denotes the generation of the physical (s)neutrinos, and the three $\lambda'$ couplings are deduced as $\lambda'^{\cal I(R)}_{vjk}\equiv\lambda'_{ijk} \tilde{\cal V}^{\cal I(R) \ast}_{vi}$, $\lambda'^{\cal N}_{vjk}\equiv\lambda'_{ijk} {\cal V}_{vi}$, and 
$\tilde{\lambda}'_{ilk}\equiv\lambda'_{ijk} K^{\ast}_{lj}$. In the following, we adopt the ``single-value-$k$'' assumption, i.e. both $\lambda'_{ij1}$ and $\lambda'_{ij2}$ are set negligible, and the NP Wilson coefficients are given at the scale $\mu_{\rm NP}=1$~TeV. 

\subsection{$B_{d(s)} \rightarrow K^{(\ast)}\bar{K}^{(\ast)}$ puzzle}
\label{sec:BKKpuzzle}
\begin{figure}[htbp]
	\centering
\includegraphics[width=0.75\textwidth]{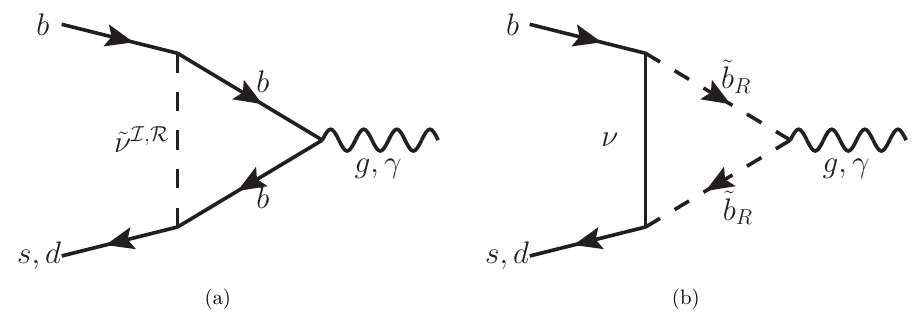}
	\caption{Gluon(photon)-penguin diagrams in RPV-MSSMIS (with $\lambda'$ couplings), within the single-value-$k$ assumption, which forbids diagrams involving charged (s)leptons with up-type squarks. 
(a): the gluon(photon)-penguin diagram involving sneutrinos. 
(b): the gluon(photon)-penguin diagram involving sbottoms.}\label{fig:bsgNP}
\end{figure}

In RPV-MSSMIS, one of the most favored operators to explain the $B_{d(s)} \rightarrow K^{(\ast)}\bar{K}^{(\ast)}$ puzzle is the magnetic one, i.e. ${\cal O}_{8gs}$, extracted from the gluon-penguins at one-loop level, shown in Fig.~\ref{fig:bsgNP}. Given that the stringent constraints from $B_s-\bar{B}_s$ mixing and $B\rightarrow X_s \gamma$ decay, which will be discussed in Sec.~\ref{sec:constraints}, are very sensitive to left-handed (LH) squark sector~\cite{Kumar:2016vhm,Hu:2019heu}, we set all LH-squarks with soft breaking masses above $10$~TeV. So the effective contributions  involve the following sparticles, i.e. sneutrinos, right-handed (RH) squarks, and gluinos, where gluinos only engage the $R$-parity conserved interaction. With the aid of the packages {\tt FeynArts}~\cite{Hahn:2000kx} and {\tt FeynCalc}~\cite{Shtabovenko:2016sxi}, all the amplitudes of these gluon(photon)-penguin diagrams can be calculated. We write the model file of RPV-MSSMIS for package {\tt SARAH}~\cite{Staub:2013tta} to generate the model file for package {\tt FeynArts}.

Next, we will show that, the NP Wilson coefficients ${C}_{7\gamma p}^{\rm NP}$ and ${C}_{8gp}^{\rm NP}$ in RPV-MSSMIS can include the chiral-flip contributions, in the mass-splitting scenario mentioned in Sec.~\ref{sec:model}. The coefficient ${C}_{8gp}^{\rm NP}$, extracted from the diagrams containing sneutrinos (other suppressed contributions omitted here, but all considered in the numerical calculations in Sec.~\ref{sec:num}) is given by,
\begin{align}\label{eq:C8g}
{C}_{8gp}^{\rm NP}=-\frac{1}{48\sqrt{2} G_F \eta_t} \biggl \{ 
& \frac{\lambda^{\prime {\cal I} \ast}_{vp3} \lambda^{\prime {\cal I} \ast}_{v33}}{m_{\tilde{\nu}^{\cal I}_v}^2}
\left[ 8 + 6 \log \left( \frac{m_b^2}{m_{\tilde{\nu}^{\cal I}_v}^2} \right) \right]
-
\frac{\lambda^{\prime {\cal R} \ast}_{vp3} \lambda^{\prime {\cal R} \ast}_{v33}}{m_{\tilde{\nu}^{\cal R}_v}^2}
\left[ 8 + 6 \log \left( \frac{m_b^2}{m_{\tilde{\nu}^{\cal R}_v}^2} \right) \right]
   \notag     \\ 
        &+
        \frac{ \lambda^{\prime {\cal I} \ast}_{vp3} \lambda^{\prime {\cal I}}_{v33} }{m^2_{\tilde{\nu}^{\cal I}_v}}
       +\frac{ \lambda^{\prime {\cal R} \ast}_{vp3} \lambda^{\prime {\cal R}}_{v33} }{m^2_{\tilde{\nu}^{\cal R}_v}}
\biggr \},         
\end{align}
and ${C}_{7\gamma p}^{\rm NP}$ is calculated as $-{C}_{8gp}/3$ because that in the setup of this model, the difference of the NP parts between $bp\gamma$ diagram and $bpg$ one is merely $-1/3$. In Eq.~\eqref{eq:C8g}, ones can find that the chiral-flip is contained in the first two terms containing double-$\lambda^{\prime\ast}$ couplings. If we utilize the degenerate-mass scenario, these two terms totally cancel with each other,  remaining the non-flip terms, ${\lambda^{\prime {\cal I} \ast}_{v23} \lambda^{\prime {\cal I}}_{v33}}/{m^2_{\tilde{\nu}^{\cal I}_v}}+{\lambda^{\prime {\cal R} \ast}_{v23} \lambda^{\prime {\cal R}}_{v33}}/{m^2_{\tilde{\nu}^{\cal R}_v}}=2{\lambda^{\prime {\cal I} \ast}_{v23} \lambda^{\prime {\cal I}}_{v33}}/{m^2_{\tilde{\nu}^{\cal I}_v}}$, which agrees with the result in Ref.~\cite{Besmer:2000rj}, with the formula-sign checked. Instead, if there exists a sufficient split between the masses $m_{\tilde{\nu}^{\cal I}_v}$ and $m_{\tilde{\nu}^{\cal R}_v}$ for the same $v$, the unique chiral-flip part is dominating, enhanced by logarithm terms. Here we analyse Fig.~\ref{fig:bsgNP}a qualitatively in the flavor basis to illustrate how double-$\lambda^{\prime\ast}$ terms are related to the chiral-flip. First the leading order term only contains normal $\lambda'\lambda^{\prime\ast}$ couplings. When we consider the next order with the mixing of chirality for one single virtual quark, the chirality of sneutrino should also be flipped, inducing that double-$\lambda^{\prime\ast}$ couplings emerge. This situation is unique since (s)neutrino chiral-flip is forbidden in original RPV-MSSM, which only contains Dirac neutrinos while no Majorana ones. 

Then it is worth mentioning that we also calculate the Wilson coefficient $C^{\rm NP}_{9\rm U}$ related to the operator, ${\cal O}_9=\frac{e^2}{16\pi^2}(\bar{s}_L \gamma^\mu b_L)(\bar{\ell}\gamma_\mu\ell)$, which  providing the lepton flavor universal contributions to $b\rightarrow s\ell^+\ell^-$, also dominated by the $\tilde{\nu}dd$ loop. The result is given that,
\begin{align}\label{eq:C9U}
C^{\rm NP}_{9\rm U}= -\frac{\sqrt{2}}{144 G_F \eta_t}\biggl \{ 
& \frac{\lambda^{\prime {\cal I} \ast}_{v23} \lambda^{\prime {\cal I}}_{v33}}{m_{\tilde{\nu}^{\cal I}_v}^2}
\left[ \frac{8}{3} + 2 \log \left( \frac{m_b^2}{m_{\tilde{\nu}^{\cal I}_v}^2} \right) \right]
+
\frac{\lambda^{\prime {\cal R} \ast}_{v23} \lambda^{\prime {\cal R}}_{v33}}{m_{\tilde{\nu}^{\cal R}_v}^2}
\left[ \frac{8}{3} + 2 \log \left( \frac{m_b^2}{m_{\tilde{\nu}^{\cal R}_v}^2} \right) \right]
   \notag     \\ 
        &+
        \frac{ \lambda^{\prime {\cal I} \ast}_{v23} \lambda^{\prime {\cal I}\ast}_{v33} }{m^2_{\tilde{\nu}^{\cal I}_v}}
       -\frac{ \lambda^{\prime {\cal R} \ast}_{v23} \lambda^{\prime {\cal R}\ast}_{v33} }{m^2_{\tilde{\nu}^{\cal R}_v}}
\biggr \},
\end{align}
and ones can find that, the chiral-flip part in Eq.~\eqref{eq:C9U}, expressed by the double-$\lambda^{\prime\ast}$ terms, is not enhanced by logarithm terms. In the degenerate-mass scenario, the result returns to the one shown in Ref.~\cite{Zheng:2021wnu}. We also examine the $Z$-penguin contribution to $C^{\rm NP}_{9\rm U}$, and find that it is negligible in the setup of this work.      

As mentioned in Sec.~\ref{sec:intro}, there are divergences between the experiment data and SM results, corresponding to the non-leptonic ratios $L_{K^{(\ast)}\bar{K}^{(\ast)}}$. If we consider NP is only within $B_s \rightarrow K^{(\ast)}\bar{K}^{(\ast)}$ decays without the $B_d$ decays,   
the predictions of the ratios, $L_{K^{(\ast)}\bar{K}^{(\ast)}}$, can be given by~\cite{Biswas:2023pyw,Lizana:2023kei},  
\begin{align}
L_{K\bar{K}}/L_{K\bar{K}}^{\rm SM}\approx& 1+1.13 C_{8gs}^{\rm NP}(\mu_{\rm EW})+0.34 C_{8gs}^{\rm NP}(\mu_{\rm EW})^2, \\
L_{K^\ast\bar{K}^\ast}/L_{K^\ast\bar{K}^\ast}^{\rm SM}\approx&1+ 2.41 C_{8gs}^{\rm NP}(\mu_{\rm EW})+1.74 C_{8gs}^{\rm NP}(\mu_{\rm EW})^2,
\end{align}
where the electroweak (EW) broken scale $\mu_{\rm EW}=160$~GeV. Then at $2\sigma$ level, we need $C_{8gs}^{\rm NP} \lesssim -0.08$ to explain the non-leptonic tension. The NP in $bd$ sector is constrained very strictly which will be shown in Sec.~\ref{sec:treeconstraint}.  

\subsection{$B \rightarrow K^{(\ast)} \nu\bar{\nu}$ revisited}
\label{sec:bsnunu}
In this section, we revisit the $B\rightarrow K^{(\ast)}\nu\bar{\nu}$ processes related to the quark transition $d_j\rightarrow d_m \nu_i\bar{\nu}_{i'}$, and the corresponding effective Lagrangian is,
\begin{align}\label{eq:Lbsnunu}
{\cal L}_{\rm eff}^{dd\nu\bar{\nu}} = &(C^{\rm SM}_{mj} \delta_{ii'} + C^{\rm NP}_{mj}) (\bar{d}_m \gamma_\mu P_L d_j)(\bar{\nu}_i \gamma^\mu P_L \nu_{i'}) + C_{mj}^{\rm 1SRR} (\bar{d}_m P_R d_j)(\bar{\nu}_i P_R \nu_{i'}) \notag\\
&+ C_{mj}^{\rm 2SRR} (\bar{d}_m \sigma_{\mu\nu} P_R d_j)(\bar{\nu}_i \sigma^{\mu\nu} P_R \nu_{i'}) + {\rm h.c.},
\end{align}
where the SM contribution is $C^{\rm SM}_{mj}=-\frac{\sqrt{2} G_F e^2 K_{tj}K^\ast_{tm}}{4\pi^2 \sin^2\theta_W} X(x_t)$ and the loop function $X(x_t) \equiv \frac{x_t(x_t+2)}{8(x_t-1)}+\frac{3 x_t (x_t-2)}{8(x_t-1)^2}\log(x_t)$ with $x_t \equiv m^2_t/m^2_W$~\cite{Buras:2014fpa}. The NP contribution of vector current is~\cite{Zheng:2022ssr}
\begin{align}\label{eq:Cbsnunu}
C^{\rm NP}_{mj} =\frac{\lambda'^{\cal N}_{i'j3}\lambda'^{\cal N\ast}_{im3}}{2 m^2_{\tilde{b}_{R}}}
= \frac{{\cal V}_{i'\alpha'} {\cal V}^{\ast}_{i\alpha} \lambda'_{\alpha'j3} \lambda'^{\ast}_{\alpha m3}}{2 m^2_{\tilde{b}_{R}}}.
\end{align}
Besides, the NP coefficients $C_{mj}^{\rm 1SRR}$ and $C_{mj}^{\rm 2SRR}$ express the chiral-flip contributions of neutrino with sbottoms. 
However, the global fit shows that, these scalar and tensor contributions are negligible, relative to the vector one~\cite{Kim:2024tsm}. For simplicity, we consider negligible mixing for the $\tilde{b}$ sector to omit these two coefficients. 

To study the NP effects on $B^+ \rightarrow K^+ \nu\bar{\nu}$ as well as $B \rightarrow K^\ast \nu \bar{\nu}$, ones can define the ratio, $R^{\nu\bar{\nu}}_{K^{(\ast)}} \equiv {{\cal B}(B \rightarrow K^{(\ast)} \nu\bar{\nu})}/{{\cal B}(B \rightarrow K^{(\ast)} \nu\bar{\nu})_{\rm SM}}$. In RPV-MSSMIS, we get that, 
\begin{align}
R^{\nu\bar{\nu}}_{K}=R^{\nu\bar{\nu}}_{K^{\ast}}
= \frac{\sum\limits_{i=1}^3 \left| C_{23}^{\rm SM} + C_{23}^{\rm NP}\right|^2 + \sum\limits_{i \neq i'}^3 \left| C_{23}^{\rm NP} \right|^2}{3 \left| C_{23}^{\rm SM} \right|^2}.
\end{align}
The recent Belle-II data~\cite{Belle-II:2023esi} of $B\rightarrow K^+ \nu\bar{\nu}$, and the updated SM prediction~\cite{Becirevic:2023aov}, induce $R^{\nu\bar{\nu}}_{K}=5.3\pm 1.7$~\cite{He:2023bnk}. Recent research~\cite{He:2023bnk} shows that, the case of $R^{\nu\bar{\nu}}_{K}=R^{\nu\bar{\nu}}_{K^{\ast}}$ cannot simultaneously fulfill the Belle-II data at $1\sigma$ level as well as the upper limit ${\cal B}(B \rightarrow K^\ast \nu\bar{\nu})_{\rm exp}<1.8 \times 10^{-5}$~\cite{Belle:2017oht}, i.e. $R^{\nu\bar{\nu}}_{K^\ast}<1.9$, at $90\%$ confidence level (CL). Besides, with the theoretical result ${\cal B}(B^+ \rightarrow K^+ \nu\bar{\nu})/{\cal B}(B \rightarrow K^\ast \nu\bar{\nu}) \approx 0.4$ for the single left-handed vector operator, it is also found that this case cannot explain the Belle-II data, without staying below the upper limit of ${\cal B}(B \rightarrow K^\ast \nu\bar{\nu})$ at $90\%$ CL~\cite{Hou:2024vyw,Chen:2024jlj}. Although $2\sigma$-level considerations for both processes may provide room for NP in this single-operator case, in this work, we still investigate the degree of $R^{\nu\bar{\nu}}_{K}$ approaching to the upper limit $1.9$, in the parameter space of RPV-MSSMIS.    

\section{The constraints}\label{sec:constraints}
Before the numerical analysis of $B_{d(s)} \rightarrow K^{(\ast)}\bar{K}^{(\ast)}$ puzzle, the relevant theoretical and experimental constraints should also be scrutinised.

\subsection{Theoretical constraints}
Firstly, some theoretical constraints on the model parameters of RPV-MSSMIS should be mentioned. As ones know, the $\lambda'$ couplings can be complex, although it may cause extra CP violations to some extent. With the consideration of perturbativity, $|\lambda'|\leqslant \sqrt{4 \pi}$ should be fulfilled~\cite{Altmannshofer:2020axr}. Moreover, the off-diagonal elements of the scalar-field trilinear couplings in the soft-breaking terms, e.g. $A_u$ in the soft-breaking term $A_u H_u \tilde{Q} \tilde{U}$, are severely restricted by the dangerous charge and colour breaking (CCB) minima and unbounded from below directions (UFB) in the effective potential
~\cite{Casas:1995pd,Casas:1996de,Casas:1997ze,Ellwanger:1999bv}. Thus, all the flavour-violating off-diagonal elements in the chiral-mixing sectors are set negligible, because the CCB and UFB bounds on them are always stronger than the ones from flavour-changing neutral-current processes, and even do not decrease when the SUSY scale increases~\cite{Casas:1995pd,Casas:1996de,Casas:1997ze}. As for the seesaw part of this model, we consider the perturbative unitarity~\cite{Chanowitz:1978mv,Durand:1989zs,Bernabeu:1993up,Fajfer:1998px,Ilakovac:1999md}, which constrains the heavy-neutrino parameter space, i.e. the mixing matrix ${\cal V}$ and heavy neutrino masses. This bound restricts the heavy neutrino decay width to comply with $\Gamma(N_v)/m_{N_v} < \frac{1}{2}$, ($v \geqslant 4$). For ${\cal V}_{vi} \sim {\cal O}(10^{-2})$, the main contribution to $\Gamma(N_v)$ is the $ N \rightarrow W l$ decay ($v \geqslant 4$), and at tree level, it is given by~\cite{Abada:2023raf},
\begin{align}
\Gamma\left(N_v \rightarrow W l_{i} \right)=
\frac{e^2}{64\pi \sin^2\theta_W} \left| {\cal V}_{v i} \right|^2 \frac{\lambda^{1/2}(m_{N_v}^2,M_W^2,m_{l_{i}}^2)}{m_{N_v}}
\left[ 1+\frac{m_{l_{i}}^2-2 M_W^2}{m_{N_v}^2}+\frac{(m_{N_v}^2-m_{l_{i}}^2)^2}{m_{N_v}^2 M_W^2} \right],
\end{align}   
where the K\"all\'en function $\lambda(a,b,c)=[a-(\sqrt{b}-\sqrt{c})^2][a-(\sqrt{b}+\sqrt{c})^2]$. The decay widths of $N_v \rightarrow Z N_{v'}$ and $N_i\rightarrow H N_j$ ($H$ is SM Higgs) are both proportional to the coupling $\Sigma_{i} {\cal V}^\ast_{v i} {\cal V}_{v i}$~\cite{Abada:2023raf}, and they are suppressed strongly.

\subsection{Direct searches} 
\label{sec:cons_direct}
Then, we move on to the experimental constraints and direct searches for SUSY particles should be considered first. Since there are no signs of NP particles until the end of the LHC run II, which reaches around $140$ fb$^{-1}$ at the center energy $13$ TeV, providing stringent bounds on SUSY models. The allowed masses of colored sparticles, such as gluinos, the first-two generation squarks, stops and sbottoms have been excluded up to $1-2$ TeV scale~\cite{Aaboud:2017opj,CMS:2018qxv,ATLAS:2019gqq,ATLAS:2020xyo,ATLAS:2021fbt,CMS:2021beq,CMS:2021eha}. In this work, the masses of LH-squarks are set above $10$ TeV, whereas the masses of sleptons as well as the heavy neutrinos are all around $10^2-10^3$~GeV. Some recent experiments have pushed the upper limit of slepton masses over TeV scale~\cite{ATLAS:2018rns,ATLAS:2018mrn,ATLAS:2021yyr}, however, these searches consider nonzero $\lambda$ related to the superpotential $\lambda_{ijk} \hat L_i \hat L_j \hat E_k$. Given that we only consider nonzero $\lambda'$ in the model, this bound can be relaxed . It is worth mentioning that, ATLAS has recently made searches for the NP signs of this type of model, only containing $\lambda'$ couplings~\cite{ATLAS:2023tlp}. Using the first collider limits for this model type, we keep $m_{\tilde{\mu}} \gtrsim 470$ GeV.

\subsection{Tree-level processes}
\label{sec:treeconstraint}
Next, we check the tree-level processes exchanging sbottoms, including $K^+ \to \pi^+ \nu \bar\nu$, $B \to \pi \nu \bar\nu$, $D^0 \to \ell^+ \ell^-$, $\tau \rightarrow \ell \rho^0$ as well as $B \to \tau \nu$, $D_s \to \tau \nu$, $\tau \to K(\pi) \nu$ and $\pi\to\ell\nu(\gamma)$.  

As ones know that the experimental measurement ${\cal B}(K^+ \to \pi^+ \nu \bar\nu)_{\rm exp}=(1.14^{+0.40}_{-0.33})\times 10^{-10}$~\cite{ParticleDataGroup:2024cfk} and the SM prediction ${\cal B}(K^+ \to \pi^+ \nu \bar\nu)_{\rm SM}=(9.24\pm0.83)\times 10^{-11}$~\cite{Aebischer:2018iyb} induce the strong constraint, $|\lambda'^{\cal N}_{i'2k}\lambda'^{\cal N\ast}_{i1k}| \lesssim 10^{-4}(m_{\tilde{b}_R}/1 {\rm TeV})^2$.
Even for ${\tilde{b}_R}$ with $10$~TeV mass, there still exists  the bound of $|\lambda'^{\cal N}_{i'2k}\lambda'^{\cal N\ast}_{i1k}| \lesssim 0.01$. Thus, we assume $\lambda'_{i1k}$ negligible to avoid this bound from this process, as well as the $B \to \pi \nu \bar\nu$ decay.

In table~\ref{tab:constraints1}, we collect the experimental results and SM predictions of $D^0 \to \ell^+ \ell^-$, $\tau \rightarrow \ell \rho^0$ decays with the charged current processes, $B \to \tau \nu$, $D_s \to \tau \nu$ and $\tau \to K \nu$, as well as the processes discussed above. Following the same/analogical numerical calculations in the ordinary RPV-MSSM (see Refs.~\cite{Earl:2018snx,Hu:2020yvs}), we update the constraint from ${\cal B}(D^0 \to \mu^+ \mu^-)$, as $|\lambda'_{223}|^2 < 0.22(m_{\tilde{b}_R}/1{\rm TeV})^2$, and the bound from ${\cal B}(D^0 \to e^+ e^-)$ is negligible due to the small $m_e$. 
We also update the calculation of the process ${\cal B}(\tau \rightarrow \ell \rho^0)$, which provides the bound $|\lambda'_{323}\lambda'^\ast_{223}| < 0.45(m_{\tilde{b}_R}/1{\rm TeV})^2$ as well as $|\lambda'_{323}\lambda'^\ast_{123}| < 0.51(m_{\tilde{b}_R}/1{\rm TeV})^2$. 
The functions $R_{133}$, $R_{223}$, and $R_{123}$ (see concrete definitions in Ref.~\cite{Hu:2020yvs}), are utilized to express the ratios of the measurement values versus the SM predictions for
${\cal B}(B \to \tau \nu)$, ${\cal B}(D_s \to \tau \nu)$, and ${\cal B}(\tau \to K \nu)$, respectively, and we also consider these constraints. 

As for the $\pi\to\ell\nu(\gamma)$ decay, similar to the formula in Ref.~\cite{Bryman:2021teu}, the bound (here also including $\lambda'$-loop corrections) can be shown with
\begin{align}\label{eq:Pidecay}
\frac{1+\eta_{\mu\mu}+h'_{\mu\mu}}{1+\eta_{ee}+h'_{ee}}=1.0010(9),
\end{align}
where the function $\eta$ and $h'$ express the non-unitary part of neutrino and $\lambda'$-loop corrections to $Wl\nu$-vertex, respectively, and they are given by ~\cite{Zheng:2022ssr}
\begin{align}
\eta_{ij} \equiv & \left({\cal V}^T_{3\times 3}\right)_{ik} {\cal U}^{-1}_{kj}-\delta_{ij},  \notag \\
h'_{li} =& -\frac{3}{64\pi^2} x_{\tilde{b}_R} f_W(x_{\tilde{b}_R}) \tilde{\lambda}'^{\ast}_{l33} \tilde{\lambda}'_{i33},
\end{align}
where ${\cal U}$ is unitary Pontecorvo–Maki–Nakagawa–Sakata (PMNS)-like, and the loop function $f_W(x) \equiv \frac{1}{x-1} + \frac{(x-2) \log x}{(x-1)^2}$ with $x_{\tilde{b}_R} \equiv m^2_t/m^2_{\tilde{b}_R}$, from the dominant $u_id_i\tilde{b}_R$-loop diagram. 
In the inverse seesaw framework, the Hermitian $\eta$ can be figured out, i.e. $\eta\approx -\frac{1}{2} m_D^\dagger (M_R^\ast)^{-1} (M_R^T)^{-1} m_D$. 
We can translate the bound Eq.~\eqref{eq:Pidecay} into $|\eta_{ee}+h'_{ee}| \lesssim 0.0028$ at the $2\sigma$ level, with the negligible $\eta(h')_{\mu\mu}$.
In this work, we can set sufficiently small $\lambda'_{2jk}$ to keep $h'_{\mu\mu}$ (negative as well as $\eta_{\mu\mu}$) negligible to avoid enlarging the Cabbibo anomaly~\cite{Coutinho:2019aiy,Blennow:2022yfm}. 

In RPV-MSSMIS, the neutrino mixing matrix, ${\cal V}$, is also bounded by the $\tau(\mu)$ decaying to charged leptons and neutrinos at the tree level. However, at one-loop level,  both ${\cal V}$ and $\lambda'$ couplings are constrained by these decays as well as the charged lepton flavor violating (cLFV) decays. We will address $\tau(\mu)$ decays totally in the following subsection~\ref{sec:loopconstraint}, and before that, we can make a summary that couplings $\lambda'_{i13}$ and $\lambda'_{2j3}$ are already set negligible (at $\mu_{\rm NP}$ scale), considering the constraints investigated above, and that is, NP is mainly not contained in the $d$ and $\mu$ sectors.   

\begin{table}[t]
\centering
\setlength\tabcolsep{8pt}
\renewcommand{\arraystretch}{1.3}
\begin{tabular}{ccc}
\hline\hline
Observations & SM predictions  & Experimental data \\
\hline
  ${\cal B}(K^+ \to \pi^+ \nu \bar\nu)$  & $(9.24\pm0.83)\times 10^{-11}$~\cite{Aebischer:2018iyb}      & $(1.14^{+0.40}_{-0.33})\times 10^{-10}$~\cite{ParticleDataGroup:2024cfk}   \\
  ${\cal B}(D^0 \to \mu^+ \mu^-)$ & $\lesssim 6\times 10^{-11}$~\cite{LHCb:2013jyo} & $<3.1\times 10^{-9}$~\cite{LHCb:2022jaa}\\
  ${\cal B}(\tau \rightarrow e \rho^0)$ & - & $<2.2\times 10^{-8}$~\cite{Belle:2023ziz}\\
  ${\cal B}(\tau \rightarrow \mu \rho^0)$ & - & $<1.7\times 10^{-8}$~\cite{Belle:2023ziz}\\
  ${\cal B}(B \to \tau \nu)$ & $(9.47\pm1.82) \times 10^{-5}$~\cite{Nandi:2016wlp} & $(1.09\pm0.24) \times 10^{-4}$~\cite{ParticleDataGroup:2024cfk} \\
  ${\cal B}(D_s \to \tau \nu)$ & $(5.40 \pm 0.30)\%$~\cite{Hu:2020yvs} & $(5.36 \pm 0.10)\%$~\cite{ParticleDataGroup:2024cfk}\\
  ${\cal B}(\tau \to K \nu)$ & $(7.15 \pm 0.026)\times 10^{-3}$~\cite{Hu:2018lmk} & 
  $(6.96 \pm 0.10)\times 10^{-3}$~\cite{ParticleDataGroup:2024cfk}\\
\hline\hline
	\end{tabular}
	\caption{Current status of the relevant processes which can be affected by RPV-MSSMIS at tree level. The experimental upper limits are given at $90\%$ CL.}
	\label{tab:constraints1}
\end{table}     

\subsection{Loop-level processes}
\label{sec:loopconstraint}
In this section illustrating loop-level bounds, we firstly investigate the $B_s-\bar{B}_s$ mixing, which is mastered by 
\begin{align}
{\cal L}_{\rm eff}^{b\bar{s}b\bar{s}}=(C^{\rm VLL}_{\rm SM}+C^{\rm VLL}_{\rm NP}) 
(\bar{s} \gamma_{\mu} P_L b)(\bar{s} \gamma^{\mu} P_L b) + 
C^{\rm 1SRR}_{\rm NP} (\bar{s} P_R b)(\bar{s} P_R b) 
+{\rm h.c.},
\end{align}
where the SM contribution is
$C_{B_s}^{\rm SM} = -\frac{1}{4 \pi^2} G_F^2 m_W^2 \eta_t^2 S(x_t)$ with the defined function $S(x_t) \equiv \frac{x_t(4-11x_t+x_t^2)}{4(x_t-1)^2}+\frac{3x_t^3\log(x_t)}{2(x_t-1)^3}$~\cite{Hu:2020yvs}. 
With the aid of {\tt FeynArts} and {\tt FeynCalc} packages, the non-negligible NP contributions are given by,
\begin{align}\label{eq:CBsMix}
C_{\rm NP}^{\rm VLL} &= \frac{1}{8i} \bigl( \frac{1}{4}
\Lambda'^{1\cal XY}_{vv'}
D_2[m_{\tilde{\nu}^{\cal X}_v},m_{\tilde{\nu}^{\cal Y}_{v'}},m_{b},m_{b}] 
+\Lambda'^{\cal N}_{vv'}
D_2[m_{\nu_v}, m_{\nu_{v'}},m_{\tilde{b}_R},m_{\tilde{b}_R}] 
\bigr), \notag\\
C_{\rm NP}^{\rm 1SRR} &= \frac{1}{8i} \bigl(
\Lambda'^{2\cal XY}_{vv'} (-1)^{\delta_{\cal XY}+1} m_b^2 D_0[m_{\tilde{\nu}^{\cal X}_v},m_{\tilde{\nu}^{\cal Y}_{v'}},m_{b},m_{b}] \notag\\
&+\Lambda'^{3\cal XY}_{vv'} (\delta_{\cal X R}-\delta_{\cal X I}) m_b^2
D_0[m_{\tilde{\nu}^{\cal X}_v},m_{\tilde{\nu}^{\cal Y}_{v'}},m_{b},m_{b}] 
-\Lambda'^{1\cal XY}_{vv'} m_b^2 D_1[m_{\tilde{\nu}^{\cal X}_v},m_{\tilde{\nu}^{\cal Y}_{v'}},m_{b},m_{b}] \notag\\
&+\Lambda'^{4\cal XY}_{vv'} (\delta_{\cal Y I}-\delta_{\cal Y R}) 
m_b^2 D_1[m_{\tilde{\nu}^{\cal X}_v},m_{\tilde{\nu}^{\cal Y}_{v'}},m_{b},m_{b}]
-\frac{\lambda'^{{\cal I} \ast 2}_{v23}}{2 m^2_{\tilde{\nu}^{\cal I}_v}}
+\frac{\lambda'^{{\cal R} \ast 2}_{v23}}{2 m^2_{\tilde{\nu}^{\cal R}_v}}
\bigr),
\end{align}
where $\Lambda'^{1\cal XY}_{vv'} \equiv \lambda'^{\cal X}_{v33} \lambda'^{\cal X \ast}_{v23} \lambda'^{\cal Y}_{v'33} \lambda'^{\cal Y \ast}_{v'23}$, 
$\Lambda'^{2\cal XY}_{vv'} \equiv \lambda'^{\cal X\ast}_{v33} \lambda'^{\cal X \ast}_{v23} \lambda'^{\cal Y \ast}_{v'33} \lambda'^{\cal Y \ast}_{v'23}$,  
$\Lambda'^{3\cal XY}_{vv'} \equiv \lambda'^{\cal X\ast}_{v33} \lambda'^{\cal X \ast}_{v23} \lambda'^{\cal Y }_{v'33} \lambda'^{\cal Y \ast}_{v'23}$, and 
$\Lambda'^{4\cal XY}_{vv'} \equiv \lambda'^{\cal X}_{v33} \lambda'^{\cal X \ast}_{v23} \lambda'^{\cal Y \ast}_{v'33} \lambda'^{\cal Y \ast}_{v'23}$
with ${\cal X}$, ${\cal Y}$ being ${\cal I}$ or ${\cal R}$, and $\Lambda'^{\cal N}_{vv'} \equiv \lambda'^{\cal N}_{v33} \lambda'^{\cal N \ast}_{v23} \lambda'^{\cal N}_{v'33} \lambda'^{\cal N \ast}_{v'23}$. The formulas of Passarino-Veltman functions~\cite{Passarino:1978jh}, $D_0$ and $D_2$, are defined as
\begin{align}
D_0[m_1,m_2,m_3,m_4] 
\equiv& \int \frac{d^4 k}{(2\pi)^4}\frac{1}{(k^2-m_1^2)(k^2-m_2^2)(k^2-m_3^2)(k^2-m_4^2)}, \notag \\
D_2[m_1,m_2,m_3,m_4] 
\equiv& \int \frac{d^4 k}{(2\pi)^4}\frac{k^2}{(k^2-m_1^2)(k^2-m_2^2)(k^2-m_3^2)(k^2-m_4^2)},
\end{align}
and $D_1$ is given by $D_\mu=p_{i\mu}D_i$~\cite{Denner:1991kt}, which is defined as
\begin{align}
D_\mu \equiv \int \frac{d^4 k}{(2\pi)^4}\frac{k_\mu}{(k^2-m_1^2)((k+p_1)^2-m_2^2)((k+p_2)^2-m_3^2)((k+p_3)^2-m_4^2)},
\end{align}
with the limit $p_i \cdot p_j \rightarrow 0$ applied.
The chiral-flip contributions are all contained in the coefficient $C_{\rm NP}^{\rm 1SRR}$, where the last two terms are extracted from the tree-level diagram. Combined with recent results of bag parameters, $B_s^{(i)}(m_b)$, including the new value of $B_s^{(1)}(m_b)$~\cite{Dowdall:2019bea,Tsang:2023nay}, we get the ratio,
\begin{align}
{\cal R}_{B_s}\equiv\frac{\Delta M_s}{\Delta M_s^{\rm SM}}=\left|1+\frac{C^{\rm VLL}_{\rm NP}}{C^{\rm VLL}_{\rm SM}}-2.38 \frac{C_{\rm NP}^{\rm 1SRR}}{C^{\rm VLL}_{\rm SM}} \right|. 
\end{align} 
The recent result averaged by the Heavy Flavor Averaging Group (HFLAV), $\Delta M_s^{\rm exp}=(17.765 \pm 0.006)$~${\rm ps}^{-1}$~\cite{HFLAV:2022esi}, 
along with the SM prediction $\Delta M_s^{\rm SM}=(18.23\pm{0.63})~{\rm ps}^{-1}$~\cite{Albrecht:2024oyn}, leads to the strong constraint $0.90 <{\cal R}_{B_s}<1.04$ at $2\sigma$ level. Given that the mass splitting of sneutrinos is considered in this work, the tree-level contributions to the ratio ${\cal R}_{B_s}$ need the cancellation to fulfill the bound.

Next we investigate the cLFV processes, i.e. $\tau \to \ell \gamma$, $\mu \to e \gamma$, $\tau \to \ell^{(')} \ell \ell$ ($\ell'\neq\ell$) and $\mu \to e e e$. It should be stressed  that the NP contributions from neutrino part, can be eliminated with the particular structures of (s)neutrino mass matrices. We utilize the structures where only chiral mixing but no flavor mixing exists for the neutrino sector involving RH neutrinos, as well as the whole sneutrino sector (see detailed discussions in Ref.~\cite{Zheng:2021wnu} and appendix~\ref{app:nmat}). Then, we focus on the $\lambda'$ contributions. 
The branching fraction of the $\tau\to\ell\gamma$ decay is given by~\cite{deGouvea:2000cf}
\begin{align}
{\cal B}(\tau\to\ell\gamma)=\frac{\tau_\tau\alpha m^5_\tau}{4}(|A^L_2|^2+|A^R_2|^2),
\end{align}
where the effective couplings $A^L_2=-{\lambda'_{\ell j3}\lambda^{\prime\ast}_{3j3}}/{64\pi^2 m^2_{\tilde{b}_R}}$ and $A^R_2=0$, with the limit of $m^2_\ell/m^2_\tau \to 0$ adopted here as well as the other cLFV processes. 
Because $\lambda'_{2jk}$ is already set negligible, processes $\mu \to e \gamma$, $\mu \to e e e$, $\tau\to\mu\gamma$, $\tau\to\mu\mu\mu$, and $\tau \to \ell^{'} \ell \ell$, will not make effective bounds. 
Then the remaining ones to be considered are $\tau\to e\gamma$ and $\tau\to eee$ decays (see the relevant formulas in Ref.~\cite{Hu:2020yvs}), with the experimental upper limits ${\cal B}(\tau\to e\gamma)_{\rm exp}<3.3\times10^{-8}$ and ${\cal B}(\tau\to eee)_{\rm exp}<2.7\times 10^{-8}$ at $90\%$ CL, respectively~\cite{ParticleDataGroup:2024cfk}. 

Following the introduction of cLFV, we will mention the $B\to X_s \gamma$ decay, which are mastered by the electromagnetic dipole $C_{7\gamma s} \approx -C_{8gs}/3$ as well as $C_{8gs}$. Although the recent SM prediction ${\cal B}(B\rightarrow X_s \gamma)_{\rm SM}\times 10^4=3.40\pm 0.17$ ($E_\gamma>1.6$~GeV)~\cite{Misiak:2020vlo}, agrees very well with the recent measured branching ratio ${\cal B}(B\rightarrow X_s \gamma)_{\rm exp}\times 10^4=3.32\pm 0.15$~\cite{Amhis:2019ckw}, which implies a very strict constraint, 
both contributions to this branching ratio from $C_{7\gamma s}$ and $C_{8gs}$ can counteract each other partly, shown as
$
{\cal B}(B\rightarrow X_s \gamma)\times 10^4=(3.40\pm 0.17)-8.25 C_{7\gamma s}(\mu_{\rm EW}) - 2.10 C_{8gs}(\mu_{\rm EW})
$~\cite{Misiak:2020vlo}, 
so RPV-MSSMIS can avoid this stringent bound, given the value of $C_{8gs}(\mu_{\rm EW})$ is expected around $-0.1$ (inducing $C_{7\gamma s}(\mu_{\rm EW})\approx 0.03$) for explaining non-leptonic puzzle. 

In the following, we move on to the purely leptonic decays of $Z$, $W$ bosons. The effective Lagrangian of $Z$-boson interaction to generic fermions $f_{i,j}$ is given by~\cite{Arnan:2019olv}
\begin{align}\label{eq:Zpole}
{\cal L}^{Zff}_{\rm eff}=\frac{e}{\cos\theta_W\sin\theta_W} \bar{f}_i \gamma^\mu \left(g_{f_L}^{ij} P_L + g_{f_R}^{ij} P_R \right) f_j Z_\mu,
\end{align}
where $g_{f_L}^{ij}=\delta^{ij}g_{f_L}^{\rm SM}+\delta g_{f_L}^{ij}$ and $g_{f_R}^{ij}=\delta^{ij}g_{f_R}^{\rm SM}+\delta g_{f_R}^{ij}$. We first investigate $Z\rightarrow l^{-}_i l^+_j$ decay, and the relevant couplings, $g_{l_L}^{\rm SM}=-\frac{1}{2}+\sin^2\theta_W$ and $g_{l_R}^{\rm SM}=\sin^2\theta_W$. 
In the limit of $m_{l_i}/m_Z \to 0$, the corresponding branching fractions are
\begin{align} 
{\cal B}(Z\to l^-_i l^+_j)=\frac{m^3_Z}{6\pi v^2 \Gamma_Z} 
\left( |g_{l_L}^{ij}|^2+|g_{l_R}^{ij}|^2 \right), 
\end{align}
with $Z$-width $\Gamma_Z=2.4955$~GeV~\cite{ParticleDataGroup:2024cfk}. For $i\neq j$, the branching ratio should be given by $[{\cal B}(Z\to l^-_i l^+_j)+{\cal B}(Z\to l^-_j l^+_i)]/2$. 
The NP effective couplings, contributed mainly by $\lambda'$ effects, are expressed as $\delta g_{l_{L}}^{ij}=\frac{1}{32\pi^2} B^{ij}$ ($\delta g_{l_{R}}^{ij}=0$) here and the formulas of $B^{ij}$ functions are given by~\cite{Earl:2018snx},
\begin{align}
B^{ij}_1 =& 3 \tilde{\lambda}'_{j33} \tilde{\lambda}'^{\ast}_{i33} \biggl\{-x_{\tilde{b}_R} (1 + \log x_{\tilde{b}_R} ) 
+ \frac{m_Z^2}{18 m^2_{\tilde{b}_R}} \biggl[(11 - 10 \sin^2\theta_W) \notag \\ 
&+ (6 - 8 \sin^2\theta_W)\log x_{\tilde{b}_R} + \frac{1}{10}(-9 + 16 \sin^2\theta_W)\frac{m_Z^2}{m_t^2} \biggr] \biggl\},   \notag\\
B^{ij}_2 =& \sum_{\ell=1}^2 \tilde{\lambda}'_{j\ell 3} \tilde{\lambda}'^{\ast}_{i\ell 3} \frac{m_Z^2}{m^2_{\tilde{b}_R}} \biggl[(1 - \frac{4}{3} \sin^2 \theta_W)(\log \frac{m_Z^2}{m^2_{\tilde{b}_R}} - i\pi -\frac{1}{3} )+
\frac{\sin^2 \theta_W}{9} \biggr].
\end{align} 
With the data of the partial width ratios of $Z$ bosons, i.e. $\Gamma(Z\to\mu\mu)/\Gamma(Z\to ee)=1.0001(24)$, $\Gamma(Z\to\tau\tau)/\Gamma(Z\to\mu\mu)=1.0010(26)$ and $\Gamma(Z\to\tau\tau)/\Gamma(Z\to ee)=1.0020(32)$~\cite{ParticleDataGroup:2024cfk}, we have the bound of $|B^{11}|<0.36$ with $|B^{33}|<0.32$. And the upper limit of the branching ratio, ${\cal}B(Z \to e\tau)<9.8\times10^{-6}$ at $95\%$ CL~\cite{ParticleDataGroup:2024cfk}, makes the bound $|B^{13}|^2+|B^{31}|^2<1.4^2$. 

Then we study the invisible $Z$-decays, i.e. $Z$ boson interaction to neutrinos, mainly in this model. The effective number of light neutrinos $N_\nu$, defined by $\Gamma_{\rm inv}=N_\nu \Gamma_{\nu\bar{\nu}}^{\rm SM}$~\cite{ALEPH:2005ab}, will constrain the relevant couplings $g_{\nu}$ in Eq.~\eqref{eq:Zpole}, via
\begin{align}
N_\nu= \sum_{i,j} \left| \delta_{ij} + \frac{\delta g_{\nu}^{ij}}{\delta g_{\nu}^{\rm SM}} \right|^2,
\end{align}   
where the coupling $\delta g_{\nu}^{\rm SM}=\frac{1}{2}$ and the formula of $\delta g_{\nu}^{ij}$ is given by~\cite{Zheng:2022irz}
\begin{align}
(32\pi^2) \delta g_{\nu}^{ij} =& \lambda'_{j33} \lambda'^{\ast}_{i33}
\frac{m^2_Z}{m^2_{\tilde{b}_R}}
\biggl\{ \left(-1+\frac{2}{3} \sin^2\theta_W \right) \left[ \log \left( \frac{m^2_Z}{m^2_{\tilde{b}_R}} \right) - i \pi - \frac{1}{3}  \right] + \left( -\frac{1}{12} + \frac{4}{9} \sin^2\theta_W \right) \biggr\}.
\end{align}
Then the measurement result $N_\nu^{\rm exp}=2.9840(82)$~\cite{ALEPH:2005ab} will make constraints.

As for the purely leptonic decays of the $W$ boson, they can be covered by the stronger ones from $\mu\to e\bar{\nu}_e\nu_\mu$ and $\tau\to\ell\bar{\nu}_{\ell}\nu_\tau$ decays. The fraction ratios of these lepton decays, i.e. ${\cal B}(\tau\to\mu\bar{\nu}_\mu\nu_\tau)/{\cal B}(\tau\to e\bar{\nu}_e\nu_\tau)$, ${\cal B}(\tau\to e\bar{\nu}_e\nu_\tau)/{\cal B}(\mu\to e\bar{\nu}_e\nu_\mu)$ and ${\cal B}(\tau\to\mu\bar{\nu}_\mu\nu_\tau)/{\cal B}(\mu\to e\bar{\nu}_e\nu_\mu)$, 
make the bounds~\cite{Bryman:2021teu} on the model parameters, which can be expressed as
\begin{align}\label{eq:cons_lepton}
\frac{1+\eta_{\mu\mu}}
{1+\eta_{ee}+ h'_{ee}}&=1.0018(14),  \notag\\
\frac{1+\eta_{\tau\tau}+ h'_{\tau\tau}}
{1+\eta_{\mu\mu}}&=1.0010(14),  \notag\\
\frac{1+\eta_{\tau\tau}+ h'_{\tau\tau}}
{1+\eta_{ee}+ h'_{ee}}&=1.0029(14).
\end{align} 
Here we only consider the $Wl\nu_l$ vertex, which has the interference with the SM contribution, and the LFV vertexes $Wl\nu_{l'}$ and $Zll'$, which can be embedded in $l\to l'\bar{\nu_i}\nu_j$ process, are omitted. With the last two formulas of Eq.~\eqref{eq:cons_lepton} combined with $|\eta_{\mu\mu}|\lesssim 10^{-4}$, we should keep 
$|\eta_{\tau\tau}+ h'_{\tau\tau}|\lesssim 0.0018$ and 
$|\eta_{\tau\tau}+ h'_{\tau\tau}|\lesssim|\eta_{ee}+h'_{ee}|$ at $2\sigma$ level.

Ones know that the $\eta$ and $h'$ functions in Eq.~\eqref{eq:cons_lepton} also affect the $W$-boson mass predictions, $m_W^{\rm NP} \approx m_W^{\rm SM} [1-0.20(\eta_{ee}+\eta_{\mu\mu}+h'_{ee})]$~\cite{Fernandez-Martinez:2016lgt}, through $G_\mu=G_F(1+\eta_{\ell\ell}+h'_{\ell\ell})$~\cite{Bryman:2021teu,Blennow:2022yfm}, where $G_\mu$ is the Fermi constant extracted from the muon lifetime, while $G_F$ corresponds to the one in the SM. Here we omit the contributions of oblique parameters, i.e. $S$, $T$, and $U$, from the self-energy of gauge bosons. In the following, we will illustrate the reason.
The one-loop contributions to the oblique parameters in RPV-MSSMIS can be divided into the MSSM part and seesaw part, and obviously, there are no $\lambda'$ couplings engaging the boson self-energy diagrams at one-loop level. The MSSM part is dominated by the one-loop diagrams involving pure squarks~\cite{Heinemeyer:2004gx} and the seesaw part is dominated by the one involving light leptons combined with heavy neutrinos~\cite{Abada:2023raf}. First we utilize {\tt SPheno}~\cite{Porod:2003um,Porod:2011nf} code, generated by package {\tt SARAH}, to calculate oblique parameters of MSSM part with inputs in table~\ref{tab:input} and the benchmark points we in table~\ref{tab:benchmark} in the following section. The results show that $S$, $T$, and $U$ from the MSSM part all stay less than $10^{-4}$  and can be omitted safely. As for the seesaw-part, the recent research shows that this bound is weaker than the one from the decays $Z\rightarrow \ell\tau$ and the invisible $Z-decays$~\cite{Abada:2023raf}, which are discussed before in this section. To summarize, the constraints from the oblique parameters can be accord with the explanation for the non-leptonic puzzle in this work.    

At last, we scrutinize all the one-loop level diagrams of $b\rightarrow s \ell^+\ell^-$ to search for the potential chiral-flip contributions. Here we use the same definitions of semi-leptonic coefficents in our recent work~\cite{Zheng:2022ssr}, i.e. $C_9^{\rm NP} = C^{\rm NP}_{\rm 9e} + C^{\rm NP}_{\rm 9U}$ and $C_{10}^{\rm NP} = C^{\rm NP}_{\rm 10e} + C^{\rm NP}_{\rm 10U}$, where $C^{\rm NP}_{\rm 9(10)e}$ express the lepton flavor non-universal contributions while $C^{\rm NP}_{\rm 9(10)U}$ express the lepton flavor universal ones. Under the ``single-value-$k$'' assumption stated in Sec.~\ref{sec:RPV-MSSMIS}, we find that there exist $C^{\rm NP}_{\rm 9e}=-C^{\rm NP}_{\rm 10e}$ as well as $C^{\rm NP}_{\rm 10U}=0$, the same as the ones of degenerate-mass scenario~\cite{Zheng:2021wnu}. 

\section{Numerical analyses}
\label{sec:num}
In this section, we begin to analyse $B_{d(s)} \rightarrow K^{(\ast)}\bar{K}^{(\ast)}$ and $B \rightarrow K^{(\ast)} \nu\bar{\nu}$ numerically within the mass-splitting scenario of RPV-MSSMIS. We consider normal ordering and $\delta_{\rm CP}=\pi$ with the recent data of neutrino oscillation~\cite{Esteban:2024eli,Esteban:2020cvm}. Then it can be calculated that the three light neutrinos have masses $\{0,0.008,0.05\}$~eV with $m_{\nu_l} \approx \{0, \sqrt{\Delta m^2_{21}}, \sqrt{\Delta m^2_{31}} \}$~\cite{Alvarado:2020lcz}. We set $\tan \beta = 5$ as the benchmark, which provides the larger room for the mass of CP-odd Higgs considering the LHC bounds~\cite{ATLAS:2019nkf,ATLAS:2020zms,ATLAS:2021upq}.  
All the sets of fixed model parameters are collected in table~\ref{tab:input}.

The diagonal inputs of $Y_\nu$, $M_R$, $m_{\tilde{L}'}$, $B_{M_R}$, and $B_{\mu_S}$ here can induce no flavor mixings in sneutrino sector, as well as the neutrino sector which RH neutrinos engage in, and this is benefit for fulfilling the bounds of cLFV decays. Besides, the input values shown in table~\ref{tab:input} can induce a diagonal $\eta=-\text{diag}(1.18,0.18,0.15)\times 10^{-3}$, which induces the $W$ mass prediction $M_W\approx 80.385$~GeV, fulfilling the recent data~\cite{ParticleDataGroup:2024cfk}. The values of $m_{\tilde{L}'_i}$ induce physical mass of the lightest sneutrino, as $m_{\tilde{\nu}_1}\approx 270$~GeV, allowed by the relevant constraints~\cite{ParticleDataGroup:2024cfk} (non-$\lambda$ and decoupled-chargino case), while the masses of charged sleptons are not affected by $B_{\mu s}$ and they are predicted as TeV scale, which in accord with the ATLAS results discussed in section~\ref{sec:cons_direct}. 
The remaining parameters, i.e. $m_{\tilde{b}_R}$, $\lambda'_{323}$, $\lambda'_{333}$, $\lambda'_{123}$ and $\lambda'_{133}$, can vary freely in the ranges considered. 

\begin{table}[t]
\centering
\setlength\tabcolsep{8pt}
\renewcommand{\arraystretch}{1.3}
\begin{tabular}{cccc}
        \hline\hline
		Parameters & Sets & Parameters  & Sets \\
		\hline
$\tan\beta$  & $5$      & $M_{\tilde{Q}_i}$   & $10$~TeV\\
$Y_\nu$  & $\text{diag}(0.28,0.11,0.10)$ & $M_{\tilde{U}_i}$     & $5$~TeV \\
$M_{R_i}$  & $1$~TeV & $m_{\tilde{L}'_i}$ & $1$~TeV   \\
$B_{M_{R_i}}$  & $0.5~\text{TeV}^2$ & $M_{\tilde{E}_i}$   & $2$~TeV \\
$B_{\mu_{S_i}}$ & $-0.66~\text{TeV}^2$ & $m_A$ & $2$ TeV \\
$A_{\nu_i}$ & $2.4$~TeV   & $A_t$ & $-4$ TeV \\
    \hline\hline
	\end{tabular}
	\caption{The sets of fixed parameters of RPV-MSSMIS.}
	\label{tab:input}
\end{table}

With the inputs given above, we can get the numerical results of the Wilson coefficient and observable, which contain chiral-flip effects, as follows,
\begin{align}\label{eq:numresults}
C^{\rm NP}_{8gs} =& (0.027 + P(m_{\tilde{b}_R}) ) \lambda^{\prime \ast}_{123} \lambda^{\prime \ast}_{133} + 0.004 \lambda^{\prime \ast}_{323} \lambda^{\prime \ast}_{333} + 0.061 \lambda^{\prime \ast}_{123} \lambda^{\prime }_{133} + 0.062 \lambda^{\prime \ast}_{323} \lambda^{\prime }_{333}, \notag \\
{\cal R}_{B_s} \approx & \left|  1 + 156.04 \lambda^{\prime \ast 2}_{123} + 21.59 \lambda^{\prime \ast 2}_{323} + B(m_{\tilde{b}_R}) \left( \lambda^{\prime \ast}_{123} \lambda^{\prime}_{133} + \lambda^{\prime \ast}_{323} \lambda^{\prime }_{333} \right)^2  \right|,   
\end{align}
where the parameters $P(m_{\tilde{b}_R})$ and $B(m_{\tilde{b}_R})$ represent the contributions from penguin diagrams involving $\tilde{b}_R$ and box ones involving both $\tilde{b}_R$ and $\tilde{\nu}$. In Eq.~\eqref{eq:numresults}, ones can see that cancellations are preferred in both the chiral-flip term $  156.04 \lambda^{\prime \ast 2}_{123} + 21.59 \lambda^{\prime \ast 2}_{323} $ and the non-flip one $\lambda^{\prime \ast}_{123} \lambda^{\prime}_{133} + \lambda^{\prime \ast}_{323} \lambda^{\prime }_{333}$, because of the stringent constraints of $B_s$-$\bar{B}_s$ mixing. The chiral-flip term demands for a tuning relation between $\lambda^{\prime}_{123}$ and $\lambda^{\prime}_{323}$, so at least, one of them should be imaginary. In the following, we set that $\lambda^{\prime}_{123}=r \lambda^{\prime}_{323} i$~ ($r=(\frac{21.59}{156.04})^{\frac{1}{2}}\approx 0.3719$ throughout this paper), and accordingly, the coupling $\lambda^{\prime}_{133}$ is also set imaginary and approaching $-i \lambda^{\prime}_{333}/r$, while couplings $\lambda^{\prime}_{323}$ and $\lambda^{\prime}_{333}$ are set real. Thus,  ones can see that $|\lambda^{\prime}_{133}|$ and $|\lambda^{\prime}_{323}|$ are the lager ones among these $|\lambda^{\prime}|$ values.

\begin{figure}[htbp]
	\centering
	\includegraphics[width=0.55\textwidth]{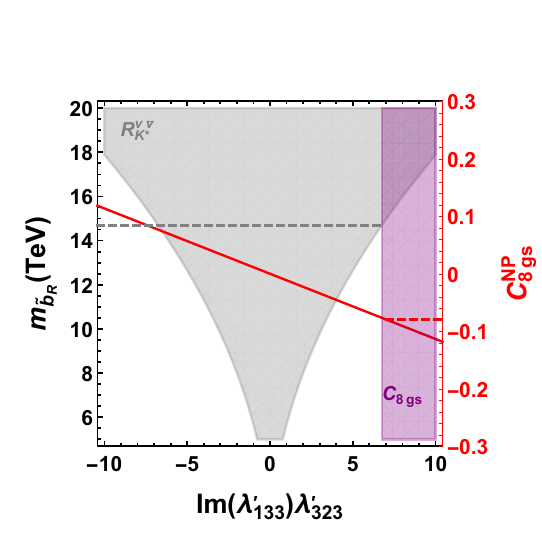}
	\caption{The $2\sigma$-level favored regions of $({\rm Im}(\lambda^{\prime}_{133}) \lambda^{\prime}_{323}, m_{\tilde{b}_R})$ for explaining the $B_{d(s)} \rightarrow K^{(\ast)}\bar{K}^{(\ast)}$ puzzle (purple region), with regions allowed by the $B\rightarrow K^\ast \nu\bar{\nu}$ data at $2\sigma$ level (gray region), combined with the Wilson coefficient $C^{\rm NP}_{8gs}$ varying with ${\rm Im}(\lambda^{\prime}_{133}) \lambda^{\prime}_{323}$, expressed by the solid red line.   
In this figure, both $\lambda^{\prime}_{123}=r \lambda^{\prime}_{323} i$ and $\lambda^{\prime}_{333} = -r {\rm Im}(\lambda^{\prime}_{133})$ are set.}
	\label{fig:analysis}
\end{figure}  

Therefore, the Wilson coefficient $C^{\rm NP}_{8gs}$, which is critical for the $B_{d(s)} \rightarrow K^{(\ast)}\bar{K}^{(\ast)}$ puzzle, is dominated by $-[ 0.01 + 0.37 P(m_{\tilde{b}_R})] {\rm Im}(\lambda^{\prime}_{133}) \lambda^{\prime}_{323}$. Given we calculate that $P(m_{\tilde{b}_R})\lesssim 10^{-3}$ for $m_{\tilde{b}_R} \gtrsim 1$~TeV, we can further get $C^{\rm NP}_{8gs} \approx -0.01 {\rm Im}(\lambda^{\prime}_{133}) \lambda^{\prime}_{323}$. 
Then, with the set, $\lambda^{\prime}_{123}=r \lambda^{\prime}_{323} i$ and $\lambda^{\prime}_{333} = -r {\rm Im}(\lambda^{\prime}_{133})$, we can get,
\begin{align}
R^{\nu\bar{\nu}}_{K^{(\ast)}} \approx & 653 X_{\tilde{b}_R}^2 
+ 0.29 \left| 1.07 - 8.78 X_{\tilde{b}_R} -(0.02+6.06 X_{\tilde{b}_R}) i \right|^2   \notag \\
&+ 0.29 \left| 1.07 - 0.44 X_{\tilde{b}_R} -(0.02+15.66 X_{\tilde{b}_R}) i \right|^2  \notag \\
&+ 0.29 \left| 1.07 + 9.18 X_{\tilde{b}_R} -(0.02-21.72 X_{\tilde{b}_R}) i \right|^2,
\end{align}
where the real-number function $X_{\tilde{b}_R}={{\rm Im}(\lambda^{\prime}_{133}) \lambda^{\prime}_{323}}/{(m_{\tilde{b}_R}/{\rm TeV})^2}$. In Fig.~2, we show the common region for non-leptonic puzzle explanation and $B \rightarrow K^\ast \nu\bar{\nu}$ data, and find that ${\rm Im}(\lambda^{\prime}_{133}) \lambda^{\prime}_{323} \gtrsim 6.8$ should be fulfilled to explain the puzzle, which induces $m_{\tilde{b}_R}$ above around $15$~TeV to below the up limit of ${\cal B}(B \rightarrow K^\ast \nu\bar{\nu})_{\rm exp}$. As shown above, even when the sbottom reach $10$~TeV scale, $R^{\nu\bar{\nu}}_{K}$ can still be close to $1.9$, provided both $|\lambda^{\prime}_{133}|$ and $|\lambda^{\prime}_{323}|$ are sufficiently large. 
However, this case will not make non-negligible effects on the $b \rightarrow s e \tau$ process, because the exchanging squark of its tree diagram is up type instead, not accord with the single-value-$k$ assumption ($\lambda^{\prime}_{i32}$ engaged). As for one-loop level, the relevant NP contributions are dominated by $4\lambda'$ boxes, given by~\cite{Zheng:2021wnu},
\begin{align}
\Delta C^{4\lambda'}_{9e\tau}=&-\Delta C^{4\lambda'}_{10e\tau}=-\frac{\sqrt{2}\pi^2 i}{2G_F \eta_t e^2} \Bigl(
\tilde{\lambda}'_{1 i3}  \tilde{\lambda}'^{\ast}_{3 i3}
\lambda'^{\cal N}_{v33} 
\lambda'^{\cal N \ast}_{v23}
D_2[m_{\nu_v},m_{u_i},m_{\tilde{b}_{R}},m_{\tilde{b}_{R}}] \notag\\
&+\tilde{\lambda}'_{1 i3}  \tilde{\lambda}'^{\ast}_{3 i3}
\lambda'^{\cal I}_{v33} 
\lambda'^{\cal I \ast}_{v23}
D_2[m_{\tilde{\nu}^{\cal I}_v},m_{\tilde{u}_{Li}},m_{b},m_{b}]
\Bigr).
\end{align}
Due to the cancellation in $\lambda^{\prime \ast}_{123} \lambda^{\prime}_{133} + \lambda^{\prime \ast}_{323} \lambda^{\prime}_{333}$, this loop contribution is also suppressed. 
Besides, as mentioned in Sec.~\ref{sec:BKKpuzzle}, there also contain chiral-flip effects in the result of $C^{\rm NP}_{\rm 9U}$, without the logarithmic enhancement. We find that $C^{\rm NP}_{\rm 9U}$ as well as $C^{\rm NP}_{\rm 9e}=-C^{\rm NP}_{\rm 10e}$, can be up to ${\cal O}(10^{-2})$ in the allowed parameter space. However, the chiral-flip contributions are negligible. 
We also check the charged processes $d_j \rightarrow u_n l \nu$, and among them, only $b \rightarrow c \tau \nu_e$ transition is affected by large $(|\lambda^{\prime}_{133}|,|\lambda^{\prime}_{323}|)$. Utilizing Eq.~(2.11) in Ref.~\cite{Zheng:2022ssr}, we find the related NP contribution versus SM one is about ${\cal O}(10^{-3})$ scale, which is negligible.  

\begin{figure}[htbp]
	\centering
	\includegraphics[width=0.95\textwidth]{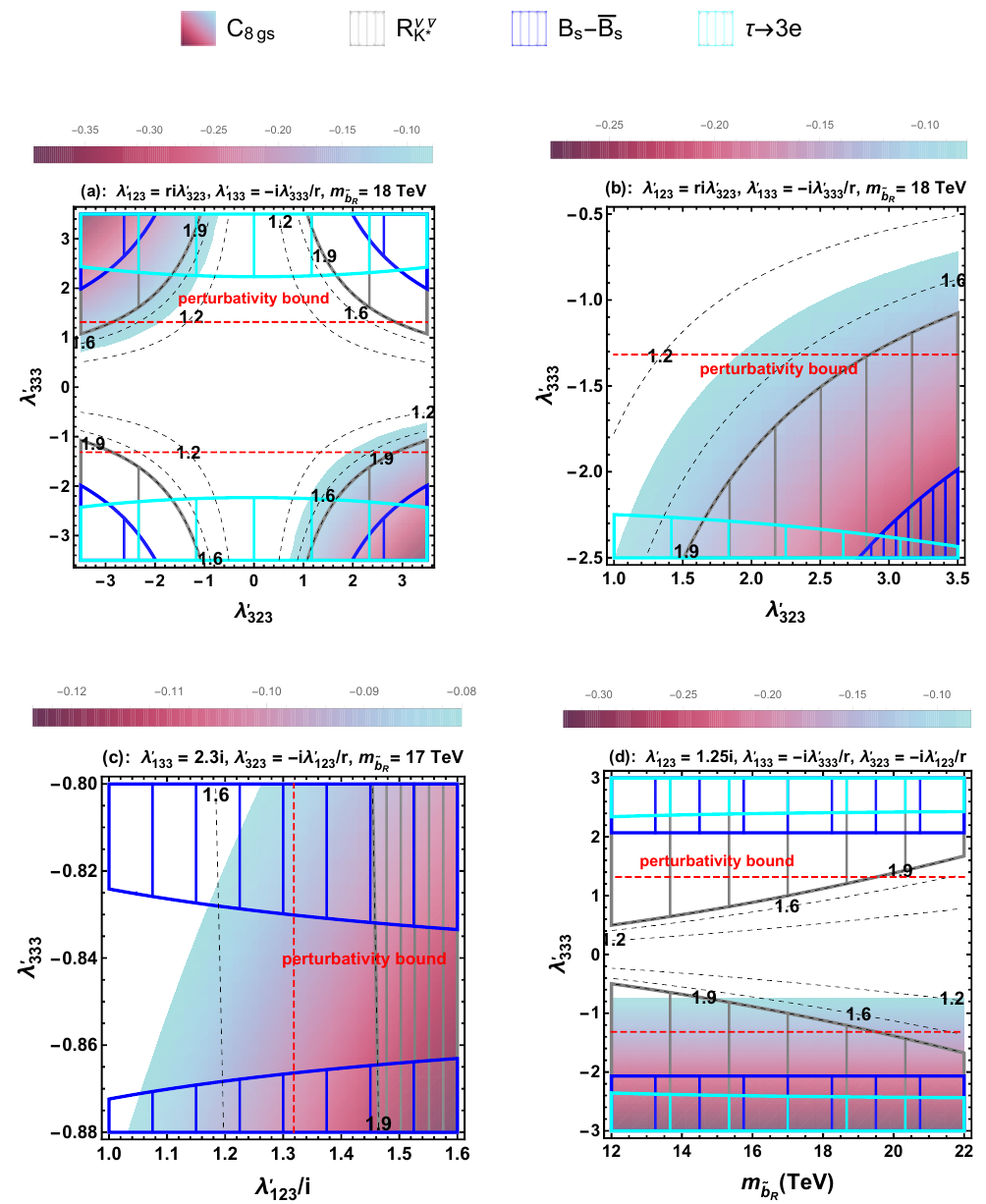}
	\caption{The $2\sigma$-level allowed regions for explaining the $B_{d(s)} \rightarrow K^{(\ast)}\bar{K}^{(\ast)}$ puzzle. The favored areas for the non-leptonic puzzle explanation is denoted by a colored gradient, where the negative coefficient $C^{\rm NP}_{8gs}$ approaches the lower (higher) value when points approaching the red (blue) area. The black dashed lines express the values of $R^{\nu\bar{\nu}}_{K^{(\ast)}}$. The hatched areas filled with the cyan, blue and gray lines are excluded by the $\tau \to e e e$ decays, $B_s-\bar{B}_s$ mixing, and $B\rightarrow K^\ast \nu\bar{\nu}$, respectively. The red dashed lines express the perturbativity limit, i.e. any $|\lambda'| \leqslant \sqrt{4\pi}$.}
	\label{fig:bound}
\end{figure}

With the rough NP features above, next, we move onto the concrete numerical analysis. As shown in Fig.~\ref{fig:bound}, ones can see that the $B_{d(s)} \rightarrow K^{(\ast)}\bar{K}^{(\ast)}$ puzzle can be explained in RPV-MSSMIS, at $2\sigma$ level. The $\tau \to e e e$ decay, $B_s-\bar{B}_s$ mixing, and $B \rightarrow K^\ast \nu\bar{\nu}$ decay, provide the dominant constraints, and the perturbativity limit is also shown. 
In Fig.~\ref{fig:bound}a, $\lambda'_{123}$ and $\lambda'_{133}$ are set related to $\lambda'_{323}$ and $\lambda'_{333}$, respectively, and $m_{\tilde{b}_R}$ is $18$ TeV. The process bounds on $\lambda'_{333}$ are mainly $\tau \to e e e$ and $B \rightarrow K^\ast \nu\bar{\nu}$ decays, while nearly overlapped by the exclusion area of perturbativity bound. 
With the same set, Fig.~\ref{fig:bound}b shows the common region in detail, which shows that $\lambda'_{323}$ should be larger than around $2.3$. And $\lambda'_{333}$ should be about lower than $-0.7$ and also higher than $-1.3$. 
In Fig.~\ref{fig:bound}c, we set $\lambda'_{133}=2.3i$, and then, the ranges for puzzle explanation are $1.02 \lesssim {\rm Im}(\lambda'_{123}) \lesssim 1.29$ and $-0.87 \lesssim \lambda'_{333} \lesssim -0.83 $. 
In Fig.~\ref{fig:bound}d, $\lambda'_{123}$ is set as $1.25i$, ones can see that the ratios $R^{\nu\bar{\nu}}_{K^{(\ast)}}$ increase with sbottom mass increasing for $\lambda'_{333}>0$, while decrease with sbottom mass increasing for $\lambda'_{333}<0$. The puzzle explanation favors $-1.29 \lesssim \lambda'_{333} \lesssim -0.8$ and $ m_{\tilde{b}_R} \gtrsim 15$~TeV. 

\begin{table}[htbp]
\centering
\setlength\tabcolsep{8pt}
\renewcommand{\arraystretch}{1.3}
\begin{tabular}{|c|c|c|c|c|c|c|c|c|c|}
\hline
$m_{\tilde{b}_R}$ & $\lambda'_{123}$ & $\lambda'_{133}$ & $\lambda'_{323}$ &  $\lambda'_{333}$ 
 & $R^{\nu\bar{\nu}}_{ K^{(\ast)}}$ & $C_{8gs}^{\rm NP}$ & $L_{K\bar{K}}$ &
$L_{K^\ast\bar{K^\ast}}$ & ${\cal B}_{VP}\times 10^{5}$  
  \\
\hline
$17$~TeV & $1.2 i$ & $2.3 i$ & $3.23$ & $-0.84$ & $1.61$ & $-0.084$ & $23.58$ &
$15.80$ & $0.80$
 \\
\hline
$19$~TeV & $1.1 i$ & $2.8 i$ & $2.96$ & $-1.04$ & $1.48$ & $-0.098$ & $23.21$ &
$15.26$ & $0.79$
 \\
\hline
$21$~TeV & $1.1 i$ & $3.0 i$ & $2.96$ & $-1.1$ & $1.37$ & $-0.102$ & $23.10$ &
$15.09$ & $0.79$
 \\
\hline
	\end{tabular}
	\caption{The benchmark points favored by the puzzle explanation. Here ${\cal B}_{VP}$ is the untagged branching ratio ${\cal B}(\bar{B}_s \rightarrow K^{\ast 0}\bar{K}^0 + c.c.)$.}
	\label{tab:benchmark}
\end{table} 

Afterwards, we collect some benchmark points in table~\ref{tab:benchmark} where the pseudoscalar-vector channel is also calculated. We consider the untagged transition $\bar{B}_s \rightarrow K^{\ast 0}\bar{K}^0$ with the branching ratio measured as ${\cal B}(\bar{B}_s \rightarrow K^{\ast 0}\bar{K}^0+c.c.)_{\rm exp}=(1.98\pm 0.28\pm 0.50)\times 10^{-5}$~\cite{LHCb:2019vww}. With the Wilson coefficient $C_{8gs}^{\rm NP}(\mu_{\rm EW})$, ones can predict this branching ratio in NP~\cite{Biswas:2024bhn},
\begin{align}
{\cal B}(\bar{B}_s \rightarrow K^{\ast 0}\bar{K}^0+c.c.)\times 10^5=0.87+0.87C_{8gs}^{\rm NP}(\mu_{\rm EW})+0.95C_{8gs}^{\rm NP}(\mu_{\rm EW})^2.
\end{align}

\section{Additional remarks}
\label{sec:addition}
Before we conclude this work, it is worth having a discussion on whether the imaginary $\lambda'$ couplings may affect CP violations. Firstly we check the NP CPV in the $B_s-\bar{B}_s$ mixing. Given the formulas of Wilson coefficients shown in Eq.~\eqref{eq:CBsMix}, along with flavor non-mixings in sneutrino content, the extra imaginary part , i.e. NP CPV not from CKM, can be only from the term, $\Lambda'^{\cal N}_{vv'}
D_2[m_{\nu_v}, m_{\nu_{v'}},m_{\tilde{b}_R},m_{\tilde{b}_R}]$, containing factor $\lambda'_{133}\lambda^{\prime\ast}_{323}{\cal V}_{v^{(\prime)}1}{\cal V}^\ast_{v^{(\prime)}3}$. However, ${\cal V}_{v^{(\prime)}1}{\cal V}^\ast_{v^{(\prime)}3}$ for light-neutrino content provides suppressing effects due to the unitarity of PMNS. In concrete numerical calculations, we confirm that this imaginary contribution can be omitted. 

Next we examine the potential CPV from $Z$ boson partical decay, which are proportional to ratios of the coupling constants, ${\rm Im}\left( {\lambda^{\prime\ast}_{iJ3} \lambda'_{iJ'3}}/{\lambda^{\prime\ast}_{1J3} \lambda'_{1J'3}}  \right)$~\cite{Barbier:2004ez}. Given we set $\lambda'_{123}$ and $\lambda'_{133}$ both purely imaginary, while $\lambda'_{323}$ and $\lambda'_{333}$ both real, these ratios have no imaginary part. 

At last, we move onto the electron electric dipole moment (EDM), related to a $u\tilde{d}\tilde{d}$-loop in the $ee\gamma$ diagram, that is proportional to the factor $[(\cos^2\beta_{\lambda'_{1jk}}-\sin^2\beta_{\lambda'_{1jk}})\sin\alpha_{A_d}+\cos\beta_{\lambda'_{1jk}}\sin\beta_{\lambda'_{1jk}}\cos\alpha_{A_d})]|\lambda'_{1jk}|^2$~\cite{Adhikari:1999pa}, where the $\alpha_{A_d}$ and $\beta_{\lambda'_{1jk}}$ are the related arguments. In the scenario of this work, we have $\beta_{\lambda'_{1jk}}=\pi/2$. With a suppressed non-positive $\alpha_{A_d}$, the constraint from electron EDM can be fulfilled. As for neutron EDMs, we should consider the $uu\gamma$ and $dd\gamma$ diagrams at one-loop level, where the $uu\gamma$ diagram always contains the $\lambda'_{i1k}$ couplings and the $dd\gamma$ diagram always  contains the $\lambda'_{i1k}$ or $\lambda'_{ij1}$ couplings. Even when  further considering two-loop contributions, e.g. Barr-Zee type, the result still contains these couplings~\cite{Chang:2000wf}. Besides, the bounds from EDMs of several nuclei and atoms, still involve $\lambda'_{ij1}$ as well as additional $\lambda'_{ij2}$~\cite{Yamanaka:2014nba}. Given that both $\lambda'_{i1k}$ and $\lambda'_{ij1(2)}$ are already set negligible in this work, the bounds of hadronic EDMs can be fulfilled safely.     

\section{Conclusions}
\label{sec:conclusion}
The recent measurements of $B_{d(s)} \rightarrow K^{(\ast)}\bar{K}^{(\ast)}$ show a non-leptonic puzzle, which expresses the deviations between the data and the QCD-factorisation prediction for the U-spin related observable, $L_{K^{(\ast)}\bar{K^{(\ast)}}}$. Besides, Belle II has recently reported the new measurement of ${\cal B}(B^+ \rightarrow K^+\nu\bar{\nu})$, around $2.7\sigma$ above the SM prediction. Both of the tensions imply that, there may exist new quark-flavor structure beyond the SM.

In this work, we study the non-leptonic puzzle and $B^+ \rightarrow K^+\nu\bar{\nu}$ in RPV-MSSMIS. This NP framework connects the trilinear interaction $\lambda'\hat L \hat Q \hat D$ with the (s)neutrino chirality flip to make the unique contribution to $L_{K^{(\ast)}\bar{K^{(\ast)}}}$, through the gluon-penguin diagrams. The chiral-flip effects are expressed as the double-$\lambda'$ terms in the Wilson coefficient $C^{\rm NP}_{8gs,d}$, which can be enhanced by the logarithm and explain the related deviation. In the $B_s-\bar{B}_s$ mixing, there also exist chiral-flip contributions, and to fulfill the strict bound of experimental data, the scenario of imaginary $\lambda'_{123}$, $\lambda'_{133}$ with real $\lambda'_{323}$, $\lambda'_{333}$ is adopted. The effect on the CPV due to this scenario is investigated as well. As for $B^+ \rightarrow K^+\nu\bar{\nu}$ decays, we find that the large $|\lambda'_{133}|$ and $|\lambda'_{323}|$, can make some enhancements, even when sbottoms are as heavy as $10$~TeV. At last, we provide some benchmark points, which also fulfill collider bounds, neutrino data, and series of flavor-physics constraints from $B$,$K$-semileptonic decays, $Z$ decays, cLFV processes, etc.

\section*{Acknowledgements}
M.D. thanks Xing-Bo Yuan for valuable discussions. This work is supported in part by Jiangxi Province Key Laboratory of Applied Optical Technology (Grant No. 2024SSY03051), the National Natural Science Foundation of China under Grant No. 12275367, the Fundamental Research Funds for the Central Universities, and the Sun Yat-Sen University Science Foundation. 

\appendix

\section{The numerical form of the (s)neutrino mixing matrix}
\label{app:nmat}
With the input set in table~\ref{tab:input}, the numerical form of the neutrino mixing matrix is listed as
\begin{align}\label{eq:VnNum}
{\cal V}^T \approx 
\left(
\begin{array}{ccccccccc}
0.842 & 0.516 & -0.152 & 0.034 i & 0 & 0 & 0.034 & 0 & 0 \\
-0.283 & 0.665 & 0.690 & 0 & 0.013 i & 0 & 0 & 0.013 & 0 \\
0.458 & -0.538 & 0.707 & 0 & 0 & 0.012 i & 0 & 0 & 0.012 \\
0 & 0 & 0 & -0.707 i & 0 & 0 & 0.707 & 0 & 0 \\
0 & 0 & 0 & 0 & -0.707 i & 0 & 0 & 0.707 & 0 \\
0 & 0 & 0 & 0 & 0 & -0.707 i & 0 & 0 & 0.707 \\
-0.040 & -0.025 & 0.007 & 0.706 i & 0 & 0 & 0.706 & 0 & 0 \\
0.005 & -0.012 & -0.013 & 0 & 0.707 i & 0 & 0 & 0.707 & 0 \\
-0.008 & 0.009 & -0.012 & 0 & 0 & 0.707 i & 0 & 0 & 0.707 \\
\end{array}
\right),
\end{align}
which is related to the neutrino mass spectrum around $\{0,8\times 10^{-15},5\times 10^{-14},1,1,1,1,1,1 \}$~TeV. And the sneutrino mixing matrices are given numerically by
\begin{align}\label{eq:VsnRNum}
{\cal \tilde{V}}^{\cal R} \approx
\left(
\begin{array}{ccccccccc}
-0.044 & 0 & 0 & -0.473 & 0 & 0 & 0.880 & 0 & 0 \\
0 & -0.018 & 0 & 0 & -0.475 & 0 & 0 & 0.880 & 0 \\
0 & 0 & -0.016 & 0 & 0 & -0.475 & 0 & 0 & 0.880 \\
0.995 & 0 & 0 & -0.100 & 0 & 0 & -0.004 & 0 & 0 \\
0 & -0.999 & 0 & 0 & 0.038 & 0 & 0 & 0.001 & 0 \\
0 & 0 & -0.999 & 0 & 0 & 0.034 & 0 & 0 & 0 \\
0.089 & 0 & 0 & 0.875 & 0 & 0 & 0.475 & 0 & 0 \\
0 & -0.034 & 0 & 0 & -0.879 & 0 & 0 & -0.475 & 0 \\
0 & 0 & -0.030 & 0 & 0 & -0.879 & 0 & 0 & -0.475 \\
\end{array}
\right),
\end{align}
related to the $m_{\tilde{\nu}^{\cal R}}$ spectrum $\{269,272,272,1010,1005,1000,1129,1127,1127 \}$~GeV, as well as
\begin{align}\label{eq:VsnINum}
{\cal \tilde{V}}^{\cal I} \approx
\left(
\begin{array}{ccccccccc}
-0.078 & 0 & 0 & -0.876 & 0 & 0 & 0.477 & 0 & 0 \\
0 & -0.032 & 0 & 0 & -0.879 & 0 & 0 & 0.475 & 0 \\
0 & 0 & -0.030 & 0 & 0 & -0.879 & 0 & 0 & 0.475 \\
-0.996 & 0 & 0 & 0.091 & 0 & 0 & 0.003 & 0 & 0 \\
0 & 0.999 & 0 & 0 & -0.037 & 0 & 0 & -0.001 & 0 \\
0 & 0 & -0.999 & 0 & 0 & 0.034 & 0 & 0 & 0 \\
0.046 & 0 & 0 & 0.475 & 0 & 0 & 0.879 & 0 & 0 \\
0 & 0.018 & 0 & 0 & 0.475 & 0 & 0 & 0.880 & 0 \\
0 & 0 & 0.016 & 0 & 0 & 0.475 & 0 & 0 & 0.880 \\
\end{array}
\right),
\end{align}
related to the $m_{\tilde{\nu}^{\cal I}}$ spectrum $\{854,854,854,1010,1005,1000,1389,1388,1388 \}$~GeV.

Then ones can find, all the chargino-sneutrino diagrams and the neutralino-slepton diagrams, among the non-$\lambda'$
diagrams in the cLFV decays of leptons, make negligible contributions due to the vanishing of flavor mixing in sneutrino sector, as shown in Eq.~\eqref{eq:VsnRNum} and Eq.~\eqref{eq:VsnINum}. As to $W/H^\pm$-neutrino diagrams, they are always connected to terms ${\cal V}^{T\ast}_{(\alpha+3)v}{\cal V}^{T}_{(\beta+3)v}$, ${\cal V}^{T\ast}_{(\alpha+3)v}{\cal V}^{T}_{\beta v}$, ${\cal V}^{T\ast}_{\alpha v}{\cal V}^{T}_{\beta v}$ and conjugate terms ($\alpha,\beta=e,\mu,\tau$ and $\alpha\neq\beta$). Readers can see calculations of these diagrams in Ref.~\cite{Abada:2014kba}. 
With the numerical form of Eq.~\eqref{eq:VnNum}, the ${\cal V}^{T\ast}_{(\alpha+3)v}{\cal V}^{T}_{(\beta+3)v}$ and ${\cal V}^{T\ast}_{(\alpha+3)v}{\cal V}^{T}_{\beta v}$ terms vanish. The ${\cal V}^{T\ast}_{\alpha v}{\cal V}^{T}_{\beta v}$ term can be decomposed into two parts, $\sum_{N=4}^{9} {\cal V}^{T\ast}_{\alpha N}{\cal V}^{T}_{\beta N}$ and $\sum_{i=1}^{3} {\cal V}^{T\ast}_{\alpha i}{\cal V}^{T}_{\beta i}=-\sum_{N=4}^{9} {\cal V}^{T\ast}_{\alpha N}{\cal V}^{T}_{\beta N}$, related to the nearly degenerate heavy neutrinos and light neutrinos respectively~\cite{Chang:2017qgi}. Then ones can also find that the ${\cal V}^{T\ast}_{\alpha v}{\cal V}^{T}_{\beta v}$ term makes no effective contribution to the cLFV decays. Thus, we conclude that the non-$\lambda'$ diagrams provide negligible effects on the cLFV decays, as mentioned in section~\ref{sec:loopconstraint}, in our input sets.

\bibliographystyle{Style1}
\bibliography{ref}

\end{document}